\documentclass[10pt]{iopart}
\usepackage{graphics,graphicx}
\usepackage{iopams}
\usepackage{cite}

\usepackage{color}

\newcommand{\ket}[1]{\mathopen|#1\mathclose\rangle}

\newcommand{\braket}[2]{\mathopen\langle#1|#2\mathclose\rangle}

\begin{document}
\title{Construction of quantum states by special superpositions of coherent states}
\author{P Adam$^{1,2}$, E Molnar$^2$, G Mogyorosi$^2$, A Varga$^2$, M Mechler$^3$ and J Janszky$^{1,2}$}
\address{$^1$Institute for Solid State Physics and Optics, Wigner Research Centre for Physics, Hungarian Academy of Sciences, H-1525 Budapest, P.O. Box 49, Hungary}
\address{$^2$Institute of Physics, University of P\'ecs, H-7624 P\'ecs, Ifj\'us\'ag \'utja 6, Hungary}
\address{$^3$MTA-PTE High-Field Terahertz Research Group, H-7624 P\'ecs, Ifj\'us\'ag \'utja 6, Hungary}
\ead{adam.peter@wigner.mta.hu}
\begin{abstract}
  We consider the optimal approximation of certain quantum states of a harmonic
  oscillator with the superposition of a finite number of coherent
  states in phase space placed either on an ellipse or on a
  certain lattice. These scenarios are currently
  experimentally feasible. The parameters of the
  ellipse and the lattice and the coefficients of the
  constituent coherent states are optimized numerically, via a genetic
  algorithm, in order to obtain the best approximation. It is found
  that for certain quantum states the obtained approximation is better
  than the ones known from the literature thus far.
\end{abstract}
\maketitle
\ioptwocol
\section{Introduction}\label{sec:intro}
There has been a persistent interest in the construction and the production of nonclassical states of a harmonic oscillator system due to their potential applications in quantum optics and quantum information processing \cite{Dodonov2002, Dodonov2007}. Special attention has also been devoted to the idea of quantum state engineering, that is, to the preparation of arbitrary quantum states in the same experimental scheme~\cite{DellAnno2006}. Several methods have been developed for generating arbitrary superpositions of photon-number states of the electromagnetic field in a cavity \cite{Vogel1993, Parkins1993, Parkins1995, Law1996}. Experimental schemes have also been proposed for the conditional preparation of arbitrary finite superposition of Fock states in a single mode of a traveling optical field by performing alternately coherent quantum-state displacement and photon addition \cite{Villas-Boas2001, Dakna1999} or subtraction \cite{Fiurasek2005}.

Another efficient method of quantum state engineering is to construct nonclassical states via discrete coherent-state superpositions.
It has been shown that superpositions of even a small number of coherent states placed along a straight line or on a circle in phase space can approximate nonclassical field states with a high degree of accuracy \cite{Janszky1993, Janszky1995, Szabo1996}.
This protocol exploits the effect of quantum interference between the constituent coherent states.
This phenomenon yields nonclassical properties of the superposition of two coherent states.
Such superpositions were originally introduced in \cite{Dodonov1974} as even and odd coherent states.
In the literature these states are generally referred as Schr\"odinger cat states due to their correspondence to Schr\"odinger's famous cat paradox.
Properties of such states have been widely discussed in the literature \cite{Dodonov2002, Dodonov2007, Buzek1992, Dodonov1995PLA, Dodonov1995PRA, Dodonov2003}, and experimental schemes have been developed for their production in different physical systems\cite{Brune1992, Brune1996, Guo1996, Auffeves2003, Jeong2006, NeegaardNielsen2006, Ourjoumtsev2007, Nielsen2007, Glancy2008, Takahashi2008, He2009, Adam2011, Lee2012, Wang2013, Laghaout2013}. 
Several methods have also been proposed for generating discrete coherent-state superpositions on a circle or along a line for electromagnetic field in cavities \cite{Szabo1996, Domokos1994p3340, Lutterbach2000, Maia2004, Pathak2005, Avelar2005, Zheng2007, Yang2010} and for the center of mass motion of a trapped ion \cite{MoyaCessa1999, Jose2000, Wang2007}.

Recently, quantum state engineering with linear optical tools has been proposed~\cite{Molnar2014}. The experimental scheme contains only beam splitters and homodyne detectors and it is capable of producing coherent state superpositions along the real axis and on an equidistant lattice in phase space. This scheme is based on an apparatus developed previously for generating Schr\"odinger cat states\cite{Adam2010}. The construction of squeezed states and number states by coherent-state superpositions on truncated von Neumann lattices was also considered in \cite{Stergioulas1999, Stergioulas1999a}. From the results in these papers one can conclude that for the considered states the accuracy of the approximation achievable using the latter lattices is below that of the superposition of the same number of coherent states placed on a circle or along a line, even after optimization by the method of \cite{Stergioulas1999}. An experimental scheme for producing coherent-state superpositions on a truncated von Neumann lattice has been developed for center of mass motion of a trapped ion \cite{AAguilar2004}. Elliptical states have also been introduced which are coherent-state superpositions on an ellipse in phase space \cite{Wang2011}. An experimental scheme was also proposed in the latter reference for generating elliptical states with constant coefficients for the motion of the center of mass of a trapped ion. Nonclassical properties of such states were also analyzed.

Inspired by these results, now we consider quantum state engineering on an ellipse and on a specific lattice in phase space. We show, by using an efficient numerical optimization method, that the superpositions of a small number of coherent states can approximate certain nonclassical states with a high accuracy. For certain states and parameter ranges the approximation is better than the corresponding one on a circle or along a line.

The paper is organized as follows. In \sref{sec:circline} we summarize the results on approximation of nonclassical states with coherent-state superpositions on a circle or along a straight line in phase space. In \sref{sec:ell} the construction of squeezed number states by coherent-state superpositions on an ellipse is considered. We present our results on approximating nonclassical states by coherent-state superpositions on a lattice in \sref{sec:latt}.

\section{Coherent state superpositions on a circle and along a straight line}\label{sec:circline}
In this section we briefly summarize the results known from the literature related to the construction of nonclassical states by coherent-state superpositions along a straight line and on a circle in phase space.
A systematic method was developed in \cite{Szabo1996} for obtaining optimized superpositions from the one-dimensional representation \cite{Janszky1990, Adam1990, Adam1990PLA, Adam1991, Buzek1991, Adam1994, Domokos1994} of the desired state.
In this quantum state engineering protocol a given quantum state $\ket{\Psi}$ is approximated either by an equidistant superposition
\begin{equation}
\ket{\psi_N}_{\rm line}=\mathcal{N}\sum_{k=1}^Nc_k^{(\rm line)}\ket{x_k}\label{eq:szabo:21}
\end{equation}
of coherent states $\ket{x_k}$ distributed at distances $d$ along the real axis of phase space, or by a superposition
\begin{equation}
\ket{\psi_N}_{\rm circle}=\mathcal{N}\sum_{k=1}^Nc_k^{(\rm circle)}\ket{R\rme^{\rmi\phi_k}}\label{eq:szabo:34}
\end{equation}
of coherent states $\ket{R\rme^{\rmi\phi_k}}$ equally distributed on an arc of a circle in phase space with radius $R$.
In \eref{eq:szabo:21} and \eref{eq:szabo:34} $\mathcal{N}$ is a normalization constant. 

The positions of the coherent states in these superpositions are defined by the following equations:
\begin{eqnarray}
x_k=x_0+\left(k-\frac{N+1}{2}\right)d,\qquad &k=1,\dots,N,\label{eq:szabo:22}\\
\phi_k=\phi_0+\left(k-\frac{N+1}{2}\right)\Delta\phi, &k=1,\dots,N,
\label{eq:szabo:35}
\end{eqnarray}
where $x_0$ and $\phi_0$ determine the center of the distribution, $\Delta\phi$ is the phase difference between two adjacent coherent states.

In the superpositions in \eref{eq:szabo:21} and \eref{eq:szabo:34}, the coefficients $c_k$ are derived from the continuous distribution functions of the one-dimensional coherent-state representations of the state $\ket{\Psi}$ along the real axis 
\begin{equation}
\ket{\Psi}=\int\limits_{-\infty}^{\infty}F(x)\ket{x}\rmd x,\label{eq:szabo:6}
\end{equation}
and on a circle
\begin{equation}
\ket{\Psi}=\int\limits_{0}^{2\pi}F_R(\phi)\ket{R\rme^{\rmi\phi}}\rmd x,\label{eq:szabo:13}
\end{equation}
respectively, by the following expressions
\begin{eqnarray}
c_k^{(\rm line)} =F(x_k),\label{eq:szabo:23}\\
c_k^{(\rm circle)} =F_R(\phi_k).\label{eq:szabo:37}
\end{eqnarray}
The optimal values of the distances $d$ and phase differences $\Delta\phi$ can be determined numerically by minimizing the misfit parameter
\begin{equation}
\epsilon=1-|\braket{\psi_N}{\Psi}|^2\label{eq:1:misfit}.
\end{equation}
The quantity $|\braket{\psi_N}{\Psi}|^2$ is the fidelity between the approximating coherent-state superposition $\ket{\psi_N}$ and the target quantum state $\ket{\Psi}$.

Obviously, the described method for deriving the approximating coherent-state superpositions can only be used if a well-behaved one-dimensional coherent-state representation  of the quantum state $\ket{\Psi}$ exists. For example, the condition of the existence of an appropriate weight function $F_R(\phi)$ on a circle with radius $R$ reads \cite{Domokos1994}
\begin{equation}
\sum_n\frac{\sqrt{n!}}{R^{n+1}}|c_n|<\infty\label{eq:szabo:18},
\end{equation}
where the numbers $c_n$ are the coefficients in the photon number expansion
\begin{equation}
\ket{\Psi}=\sum_{n=0}^{\infty}c_n\ket{n}.\label{eq:szabo:1}
\end{equation}
We note that for a state with an infinite number of Fock-state coefficients not fulfilling \eref{eq:szabo:18} an approximate weight function $F^{\rm appr}_R(\phi)$ can be derived by appropriate truncation of the expansion of \eref{eq:szabo:1}. As a consequence, discrete coherent-state superpositions approximating a nonclassical state can always be constructed on a circle in phase space.

Using this procedure, approximating discrete coherent-state superpositions were determined in \cite{Szabo1996} for various nonclassical states, including displaced squeezed number states, squeezed coherent states, binomial states, Hermite-polynomial states, and amplitude squeezed states. In \cite{Szabo2000} coherent-state superpositions approximating phase-optimized states on the real axis of  phase space were derived by a different numerical method. This method varies not only the coefficients but also the positions of the coherent states in phase space. Though there is no one-dimensional straight-line coherent-state representation known for these states, the numerically obtained superpositions of high accuracy consist of a few coherent states distributed nearly equidistantly. 
This experience can be explained by the properties of the quantum interference between the constituent coherent states.
The interference phenomenon arises typically at a certain range of distance between coherent states.
In view of this, the choice of the equidistant states in the superpositions \eref{eq:szabo:21} and \eref{eq:szabo:34} seems to be reasonable. 
We note that the accuracy of the approximation can be generally increased by increasing the number of constituent states in the superpositions.

From these results one can conclude that any state can be approximated by discrete coherent-state superpositions.
Even expecting a given accuracy, different superpositions of various geometries can approximate the same nonclassical state.
Coherent state superpositions along a straight line and on a circle have special importance because such superpositions can be generated experimentally in various physical systems \cite{Szabo1996, Domokos1994p3340, Lutterbach2000, Maia2004, Pathak2005, Avelar2005, Zheng2007, Yang2010,MoyaCessa1999, Jose2000, Wang2007}.
In general, small number of constituent states are required if the geometry of their positions fits well to the state to be approximated.
For example, states exhibiting circular symmetry in the Wigner function such as photon number states or amplitude squeezed states can be effectively approximated by superpositions on a circle.
The optimal value of the radius of the circle fits to the value of the photon number \cite{Szabo1996} when the constituent coherent states in the superposition  are located under the central part of the Wigner function of the target state.
On the other hand, effective approximation of quadrature squeezed states can be realized by superpositions along a line.
These facts motivate our considerations of coherent-state superpositions on an ellipse for approximating squeezed number states.

\section{Coherent-state superpositions on an ellipse}\label{sec:ell}
Elliptical states has been introduced in \cite{Wang2011}. These states can be defined as coherent-state superpositions on an ellipse in phase space
\begin{equation}
	\ket{\psi_N}_{\rm{ellipse}} = \mathcal{N}\sum_{k=1}^N c_k^{(\rm{ellipse})} \ket{\alpha_k}, \label{eq:state:ell}
\end{equation}
where
\begin{eqnarray}
		\alpha_k = r_k\exp(\rmi\phi_k),\label{eq:ell:alpha}\\
		\phi_k = \phi_0+\frac{2\pi k}{N},\label{eq:ell:phi}\\
		r_k = \left( \frac{\cos^2\phi_k}{a^2} + \frac{\sin^2\phi_k}{b^2}\right)^{-1/2},\label{eq:ell:r}
\end{eqnarray}
where $a$ and $b$ are arbitrary real numbers. Here the constituent coherent states are placed equidistantly in their phase $\phi_k$. Nonclassical properties of such states were analyzed only for states with constant coefficients $c_k=1$ \cite{Wang2011}.

Here we consider the construction of quantum states via elliptical states. Inspired by the considerations in the previous section, we choose displaced squeezed number states for demonstrating the idea of approximating quantum states by discrete coherent-state superpositions on an ellipse.

Displaced squeezed states are defined as
\begin{equation}
\ket{n,\zeta,Z}=\hat{D}(Z)\hat{S}(\zeta)\ket{n}\label{eq:3:sdstate},
\end{equation}
where $\hat{D}(Z)$ is the displacement operator
\begin{equation}
\hat{D}(Z)=\exp(Z\hat{a}^{\dagger}-Z^*\hat{a}),
\end{equation}
while $\hat{S}(\zeta)$ is the squeezing operator 
\begin{equation}
\hat{S}(\zeta)=\exp\left(\frac{1}{2}\zeta^*\hat{a}^2-\frac{1}{2} \zeta\hat{a}^{\dagger 2}\right), \qquad \zeta=r\rme^{\rmi\theta}.
\end{equation}
In the above definitions $\hat{a}$ and $\hat{a}^\dagger$ are the annihilation and creation operators, respectively, $Z$ is the amount of displacement in phase space, and $\zeta$ is the complex squeezing parameter.

The Wigner function for a squeezed number state reads
\begin{eqnarray}
W_{\rm SN}&(\alpha)=\frac{2}{\pi}\exp\left[\frac{1}{2}(\alpha-\alpha^*)^2\rme^{-2r}-\frac{1}{2}(\alpha+\alpha^*)^2\rme^{2r}\right]\nonumber\\
&\times(-1)^n\mathcal{L}_n\left[(\alpha+\alpha^*)^2\rme^{2r}-(\alpha-\alpha^*)^2\rme^{-2r}\right]\label{eq:Wigner:1},
\end{eqnarray}
where $\mathcal{L}_n(x)$ is the Laguerre polynomial of degree $n$:
\begin{equation}
\mathcal{L}_n(x)=\sum_{m=0}^n(-1)^m{n \choose m}\frac{x^m}{m!}.
\end{equation}

\begin{figure}
\centering
\includegraphics[width=0.8\columnwidth]{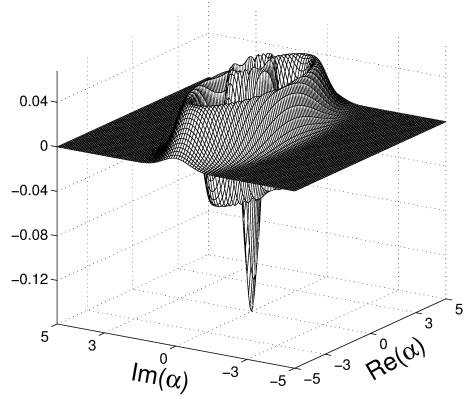}
\caption{\label{fig:sn305css}Wigner function $W_{\rm SN}(\alpha)$ for the squeezed number state $\ket{3,0.5,0}$.}
\end{figure}

\begin{figure}
\centering
\includegraphics[width=0.8\columnwidth]{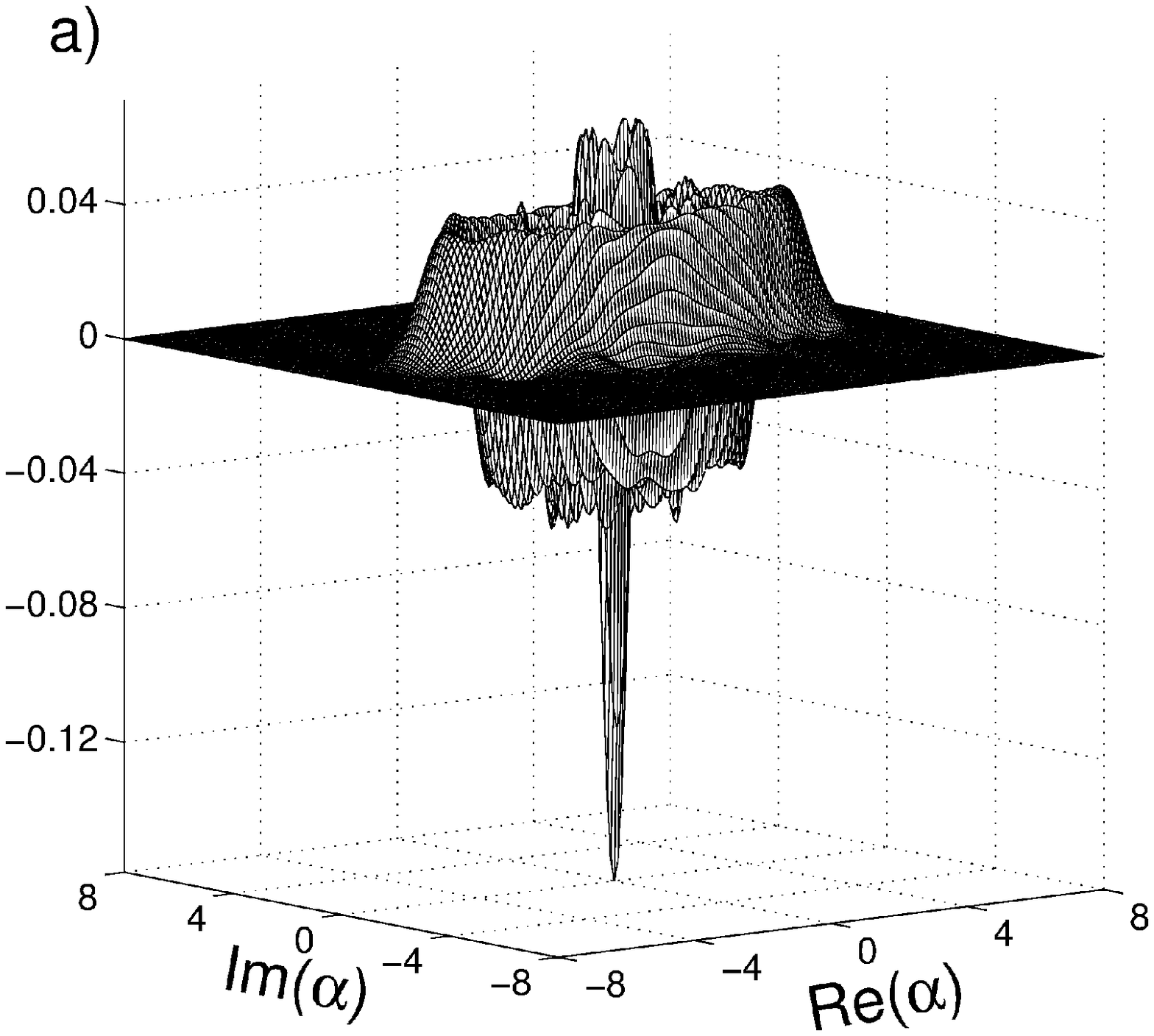}
\includegraphics[width=0.8\columnwidth]{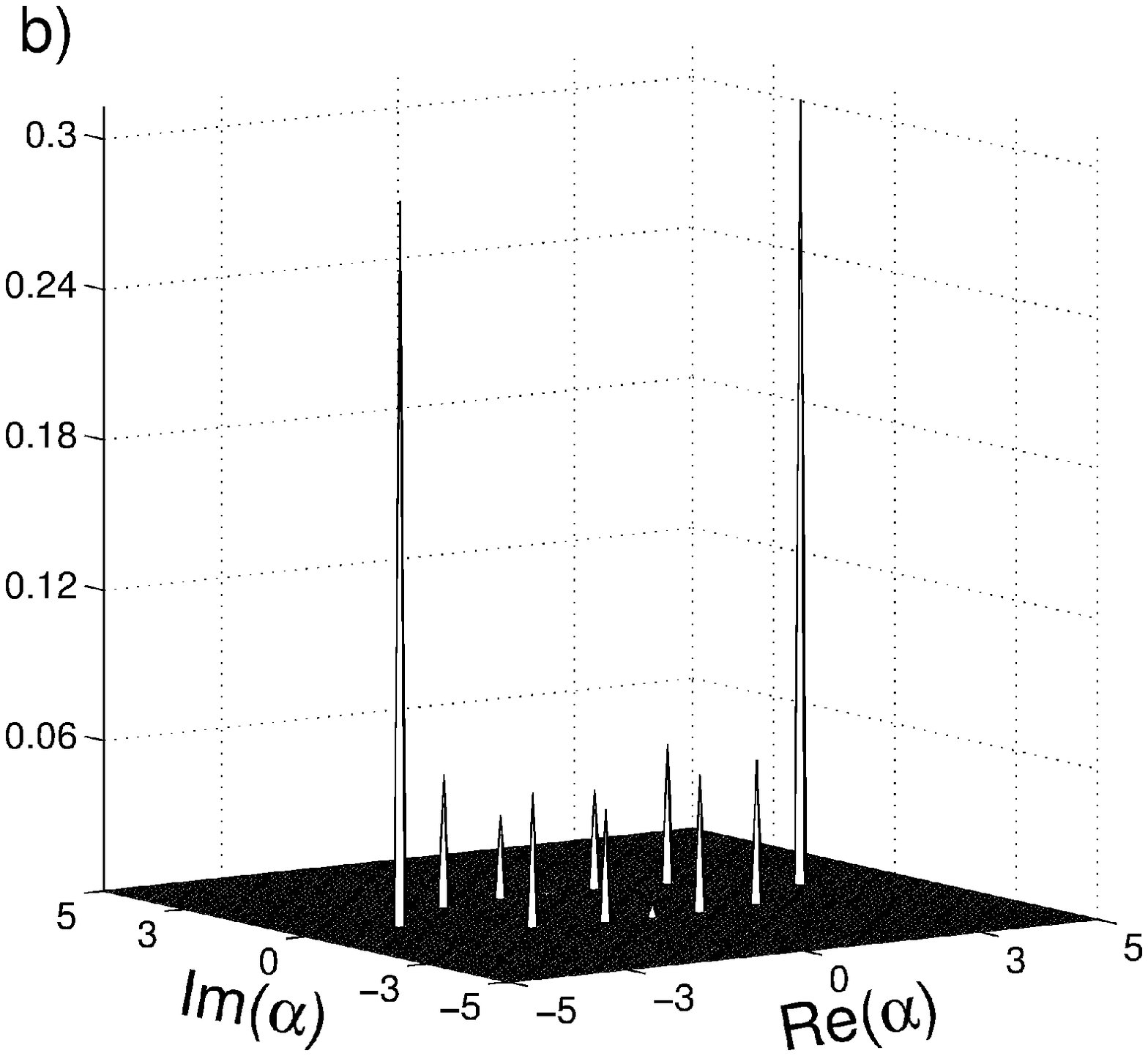}
\caption{\label{fig:505ell}Squeezed number state $\ket{5,0.5,0}$ approximated by $N=12$ coherent states on a ellipse (a) Wigner function $W(\alpha)$ of the approximating coherent-state superposition, (b) pins showing the positions and the absolute values of the coefficients of the constituent coherent states.} 
\end{figure}

In \fref{fig:sn305css} the Wigner function $W(\alpha)$ of the state $\ket{3,0.5,0}$ is presented.
It can be seen that the Wigner function of the squeezed number states has an elliptic shape. We note that for the displaced number state the one-dimensional coherent-state distribution function is also known, so approximating discrete coherent-state superpositions along the real axis can be easily obtained using the method described in \sref{sec:circline}.

Our task is to find the set of coefficients $c_k^{(\rm ellipse)}$ in \eref{eq:state:ell} and parameters $a$, $b$ of the ellipse in \eref{eq:ell:r}, which minimize the misfit parameter in \eref{eq:1:misfit}. Since it is not a convex function of the parameters, an algorithm supporting non-convex optimization has to be chosen. 
In our calculations we have chosen to use a genetic algorithm~\cite{Goldberg} for solving the optimization problem, as its applicability is quite general and we found it effective in the present case. 
In order to eliminate the uncertainty of the method due to its stochastic nature, we have performed multiple runs and we chose the parameter set for which the value of the misfit parameter \eref{eq:1:misfit} was minimal. Besides, at a given accuracy, various solutions with slightly different parameters can be found for a given target state. This degeneracy of the optimum can be foreseen from the general properties of the discrete coherent-state superposition described in the last paragraph of \sref{sec:circline}.

In tables~\ref{tab:ell1}-\ref{tab:ell4} we present our results on approximating displaced squeezed number states on an ellipse with 12 coherent states for various values of the photon number, the squeezing parameter and the amount of displacement. We show the misfit $\epsilon^{(\rm ellipse)}$ defined in \eref{eq:1:misfit} and the ellipse parameters $a_{\rm opt}$ and $b_{\rm opt}$ introduced in \eref{eq:ell:r} corresponding to the optimum. For comparison, the tables also show the misfit $\epsilon^{(\rm line)}$ and the optimal distance $d_{\rm opt}$ for approximating displaced squeezed number states along the real axis of  phase space with 12 equidistant coherent states. We note that we determined the coefficients $c_k^{(\rm line)}$ and the optimal distances $d_{\rm opt}$ of the constituent coherent states in the line superpositions by using the same genetic algorithm. This method may lead to a better approximation than the one described in \sref{sec:circline}.

\begin{table}
\caption{\label{tab:ell1}The minimal misfits $\epsilon^{(\rm ellipse)}$ and $\epsilon^{(\rm line)}$ of the approximation  of the squeezed number states on an ellipse and along a line with 12 coherent states and the optimal parameters $a_{\rm opt}$, $b_{\rm opt}$ of the ellipse and $d_{\rm opt}$ of the line for various photon numbers and for a constant squeezing parameter $\zeta=0.5$.}
\begin{indented}
\item[]\begin{tabular}{llllll}
		\br
			State & $\epsilon^{(\rm ellipse)}$   & $a_{\rm opt}$ & $b_{\rm opt}$&$\epsilon^{(\rm line)}$ & $d_{\rm opt}$\\
		\mr
			$\ket{0,0.5,0}$ & $6\times10^{-4}$& 1.26 & 0.17& $1.4 \times 10^{-5}$ & 0.27\\ 
			$\ket{3,0.5,0}$ & 0.001 & 2.63 & 0.97&0.0015 & 0.47\\
			$\ket{5,0.5,0}$ &	0.006	& 3.39 & 1.44&0.018 & 0.6\\
			$\ket{7,0.5,0}$ &	0.0172 &	4.09	&1.54&0.052 & 0.74\\
		\br
	\end{tabular}
\end{indented}
\end{table}
In \tref{tab:ell1} our results are presented for a fixed real squeezing parameter $\zeta=0.5$ and no displacement, while the photon number $n$ is changed. It can be seen that increasing the photon number leads to the growth of the size of the optimal ellipse while the accuracy of the approximation $\epsilon$ slightly decreases but remains rather high for all the states considered. According to the table the approximation on an ellipse proves to be better than approximation along a line except for the squeezed vacuum state $\ket{0,0.5,0}$. This effect confirms the intuition described at the end of the previous section that the best performance can be achieved by fitting the location of the constituent coherent states in the superposition to the geometry and the size of the Wigner function of the target state.
\Fref{fig:505ell} shows the pins representing the positions and the absolute values of the coefficients of the $N=12$ constituent coherent states in the coherent-state superposition approximating the squeezed number state $\ket{5,0.5,0}$  on an ellipse. The Wigner function $W(\alpha)$ of the approximating superposition is also shown. This function practically coincides with the Wigner function of the target state, only some imperfections can be seen appearing in the form of small fluctuations due to quantum interference.

\begin{table}
\caption{\label{tab:ell2}The misfits $\epsilon^{(\rm ellipse)}$ and $\epsilon^{(\rm line)}$ of the approximation  of the squeezed number states on an ellipse and along a line with 12 coherent states and the optimal parameters $a_{\rm opt}$, $b_{\rm opt}$ of the ellipse and $d_{\rm opt}$ of the line for various squeezing parameters and for a constant photon number $n=3$.}
\begin{indented}
\item[]\begin{tabular}{llllll}
		\br
			State & $\epsilon^{(\rm ellipse)}$   & $a_{\rm opt}$ & $b_{\rm opt}$&$\epsilon^{(\rm line)}$ & $d_{\rm opt}$\\
		\mr
$\ket{3,0.1,0}$&$5 \times 10^{-5} $&1.48 &1.56&0.0099& 0.3\\ 
$\ket{3,0.3,0}$&$2.7\times10^{-4}$&2.07 &0.89&0.0035& 0.41\\ 
$\ket{3,0.5,0}$ & 0.001& 2.63 & 0.97&0.0015 & 0.53\\
$\ket{3,0.7,0}$ &	0.011& 3.12 & 0.94&0.0033 & 0.71\\
$\ket{3,0.8,0}$ &	0.017& 3.55 & 0.99&0.0028 & 0.76\\
$\ket{3,1.2,0}$ &	0.12 &	4.91	& 0.75&0.0086 & 1.16\\
		\br
	\end{tabular}
\end{indented}
\end{table}
\Tref{tab:ell2} contains results for constant photon number $n=3$  without displacement and for various real squeezing parameters $\zeta$. It can be seen from the table that for stronger squeezing the optimal ellipse becomes more and more elongated, that is, the ratio $a/b$ of the parameters of the ellipse increases, in the same way as the Wigner function of the state becomes more elongated. The data show that the superposition on an ellipse proves to be better only for smaller squeezing parameters. For a stronger squeezing when the Wigner function of the state is dominantly elongated along the real axis, superpositions along the appropriate line perform better.

\begin{table}
\caption{\label{tab:ell3}The misfits $\epsilon^{(\rm ellipse)}$ and $\epsilon^{(\rm line)}$ of the approximation  of the displaced squeezed number states on an ellipse and along a line with 12 coherent states and the optimal parameters $a_{\rm opt}$, $b_{\rm opt}$ of the ellipse and $d_{\rm opt}$ of the line for a constant squeezing parameter $\zeta=0.5$, a constant photon number $n=3$, and for various displacements $Z$.}
\begin{indented}
\item[]\begin{tabular}{llllll}
		\br
			State & $\epsilon^{(\rm ellipse)}$   & $a_{\rm opt}$ & $b_{\rm opt}$&$\epsilon^{(\rm line)}$ & $d_{\rm opt}$\\
		\mr
			$\ket{3,0.5,0.2\rmi}$ & 0.0041 & 2.62 &1& 0.0078 & 0.46\\	
			$\ket{3,0.5,0.3\rmi}$ & 0.0035 & 2.64 &1.33& 0.0046 & 0.67\\		
			$\ket{3,0.5,0.5\rmi}$ & 0.009 & 2.49 &1.6& 0.028 & 0.7\\
			$\ket{3,0.5,0.8\rmi}$ & 0.038&2.37& 1.73&0.12 & 0.62\\ 
			$\ket{3,0.5,\rmi}$ & 0.088&2.58& 1.72& 0.29 & 0.53\\
			$\ket{3,0.5,1.5\rmi}$ & 0.026& 2.87 &1.94 & 0.47 & 0.5 \\
			$\ket{3,0.5,1.8\rmi}$ & 0.068& 2.63 &2.52 & 0.56 & 0.55 \\
			$\ket{3,0.5,2\rmi}$ &	 0.12 &	2.5	& 3.41&0.65 & 0.24\\
		\br
	\end{tabular}
\end{indented}	
\end{table}
The results for displaced squeezed number states are shown in \tref{tab:ell3}. In this case both the photon number $n=3$ and the squeezing parameter $\zeta=0.5$ are kept constant, while the amount of displacement $Z$ is changed. According to the table, increasing the displacement results generally in a decrease of the accuracy, but exceptions can occur at certain values of the displacement. Though the tendency for approximating along a line is the same, the approximation is much less accurate. The data also show that the optimal ellipse tends to deform towards the displacement ensuring the interfering coherent states to be close to the area of the Wigner function of the displaced squeezed target state. Interestingly, the optimal length $a_{\rm opt}$ of  the major axis of the ellipse does not change monotonously.
\begin{figure}
\centering
\includegraphics[width=0.8\columnwidth]{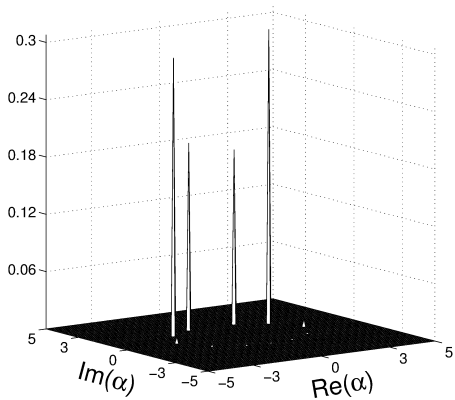}
\caption{\label{fig:tuskeabra}Pins showing the positions and the absolute values of the coefficients of the constituent coherent states for approximating squeezed displaced number state $\ket{3,0.5,1.5\rmi}$ with $N=9$ coherent states on a lattice.}
\end{figure}
From \fref{fig:tuskeabra} it can be seen that for a larger displacement $Z=1.5\rmi$ only superposed coherent states in the upper half of phase space have non-negligible weights.

\begin{table}
\caption{\label{tab:ell4}The misfits $\epsilon^{(\rm ellipse)}$ and $\epsilon^{(\rm line)}$ of the approximation  of the squeezed number states on an ellipse and along a line with 12 coherent states and the optimal parameters $a_{\rm opt}$, $b_{\rm opt}$ of the ellipse and $d_{\rm opt}$ of the line for various phase $\theta$ and a constant magnitude $r=0.5$ of squeezing and a constant photon number $n=3$.}
\begin{indented}
\item[]\begin{tabular}{l@{\;}lllll}
		\br
			State & $\epsilon^{(\rm ellipse)}$   & $a_{\rm opt}$ & $b_{\rm opt}$&$\epsilon^{(\rm line)}$ & $d_{\rm opt}$ \\
		\mr
			$\ket{3,0.5,0}$ & 0.001 & 2.63 & 0.97&0.0015&0.53 \\
			$\ket{3,0.5\exp(\rmi\pi/12),0}$ & 0.006 & 2.77 &1.33& 0.0018 & 0.52\\ 
			$\ket{3,0.5\exp(\rmi\pi/6),0}$ & 0.017 &2.62 &1.61& 0.012 & 0.47\\
			$\ket{3,0.5\exp(\rmi\pi/4),0}$ &	0.03	& 2.37 &1.6&0.09 & 0.46\\
			$\ket{3,0.5\exp(\rmi\pi/2),0}$ & 0.022 &2.06 &2.01& 0.29 & 0.4\\
			$\ket{3,0.5\exp(\rmi5\pi/6),0}$ & 0.0173 &1.64 &2.6& 0.59 & 0.29\\
			$\ket{3,0.5\exp(\rmi11\pi/12),0}$ &	0.0067	& 1.42 &2.66 & 0.55 & 0.24 \\ 
			$\ket{3,0.5\exp(\rmi \pi),0}$ &0.0011& 0.97&2.59&0.6 & 0.28\\
		\br
	\end{tabular}
\end{indented}
\end{table}
Finally, \tref{tab:ell4} presents results on squeezed number states with various phases $\theta$ but constant magnitude $r=0.5$ of squeezing, keeping the photon number $n=3$ fixed. From the data it can be seen that the deformation of the ellipse changes in accordance with the increase of the phase of squeezing. The originally minor axis $b$ of the ellipse is growing, for phases larger than $\pi/2$ the ellipse elongates in the direction of the imaginary axis of phase space. The change in accuracy is not monotonous, the accuracy is the highest for squeezing corresponding to the orientation of the ellipse, that is, for $\theta=0$ and $\theta=\pi$. For larger phases of squeezing ($\theta>\pi/6$) approximation on an ellipse proves to be better than the one along a line.

\section{Discrete coherent-state superpositions on a lattice in phase space \label{sec:latt}}
Recently, it has been shown that discrete coherent-state superpositions with variable coefficients on a lattice in phase space can be produced in traveling wave optics using only beam splitters and homodyne measurements \cite{Molnar2014}.
In this section we consider the approximation of nonclassical states by such \emph{experimentally realizable} superpositions on a lattice in phase space.

Accordingly, let us consider the superposition of 9 coherent states
\begin{equation}
\ket{\psi_9}_{\rm lattice} = \mathcal{N}\sum_{l=-1}^1 \sum_{k=-1}^1 c_{k,l}^{(\rm lattice)} \ket{l \cdot d + k \cdot \rmi d}, \label{eq:dcss}
\end{equation}
on an equidistant lattice centered around the origin in phase space. In this equation $d$ is the distance between adjacent elements of the lattice and $\mathcal{N}$ is a normalization constant.

Again, we have used the same genetic algorithm as in the previous section for finding the optimal complex coefficients $c_{k,l}^{(\rm lattice)}$ and the distance $d$ in \eref{eq:dcss}. For the measure of the accuracy of the approximation to be optimized, we use the misfit parameter $\epsilon^{(\rm lattice)}$ introduced in \eref{eq:1:misfit}. For comparison, we also determined the misfit $\epsilon^{(\rm circle)}$ and the optimal radius $R_{\rm opt}$ for equidistant coherent-state superpositions consisting 9 coherent states on a circle and approximating the given target state.

\begin{figure}
\centering
\includegraphics[width=0.8\columnwidth]{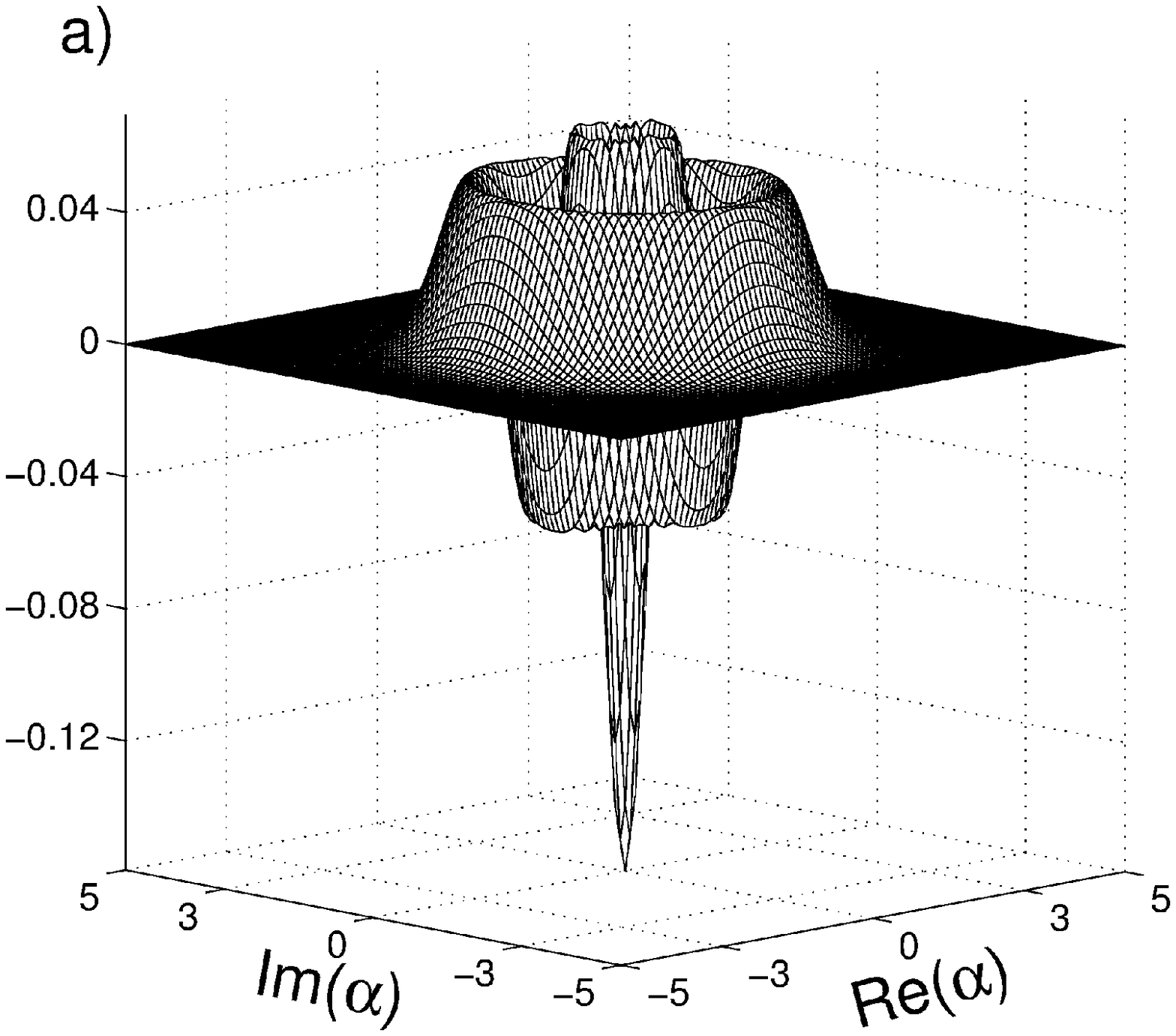}
\includegraphics[width=0.8\columnwidth]{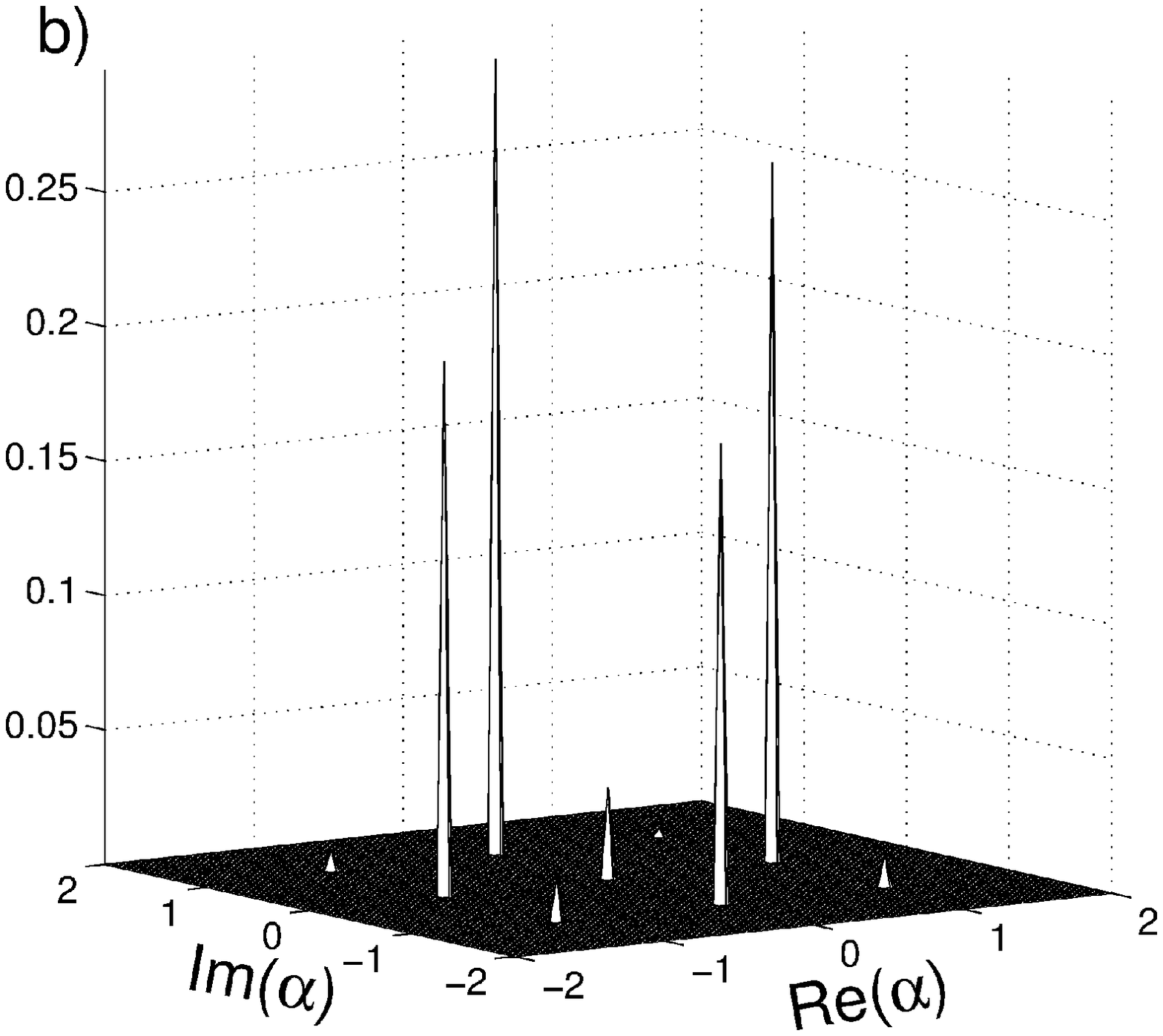}
\caption{\label{fig:number5}Number state $\ket{3} $ approximated by $N=9$ coherent states on a lattice (a) Wigner function $W(\alpha)$ of the approximating coherent-state superposition, (b) pins showing the positions and the absolute values of the coefficients of the constituent coherent states.}
\end{figure}

First we examine the problem of constructing number states by lattice and circle superpositions. The results are presented in \tref{tab:number}.
\begin{table}
\caption{\label{tab:number}The misfits $\epsilon^{(\rm lattice)}$ and $\epsilon^{(\rm circle)}$ of the approximation  of the number states on a lattice and on a circle with 9 coherent states and the optimal distances $d_{\rm opt}$ and radii $R_{\rm opt}$ for various photon numbers $n$.}
\begin{indented}
\item[]\begin{tabular}{lllll}
		\br
			State & $\epsilon^{(\rm lattice)}$ &$d_{\rm opt}$ & $\epsilon^{(\rm circle)}$ & $R_{\rm opt}$\\
		\mr
			$\ket{1}$&$2.2\times10^{-6}$&0.22 & $3.4\times10^{-6}$ & 0.23 \\
			$\ket{2}$&$3.1\times10^{-4}$&0.39 & $5.6\times10^{-6}$ & 0.3\\
			$\ket{3} $ &$3.5\times10^{-4}$& 0.5 & $3.9\times10^{-5}$ & 0.55\\
			$\ket{4} $ &$5\times10^{-4}$&0.73 &$1.3\times10^{-5}$ & 0.8\\
			$\ket{5}$ & 0.004 & 1.03 &$2.8\times10^{-4}$& 1.25\\
			$\ket{6} $ & 0.0085 &1.2  & $4.8\times10^{-5}$ & 1.73\\
			$\ket{7}$ &  0.009 & 1.35 &$8.3\times10^{-5}$ & 1.92\\
			$\ket{8}$&0.0097&1.5 &$2.1\times10^{-4}$& 1.99\\
		\br
		\end{tabular}
\end{indented}
\end{table}
It can be seen that all the considered number states ($n\leq 8$) can be approximated with high accuracy on both formations. The data show that the approximation on a circle is better than the corresponding one on a lattice for number states $\ket{n}$ except for $n=1$. The optimal distance $d_{\rm opt}$ and radius $R_{\rm opt}$ increases with  $n$, that is, with the growth of the area of the Wigner function of the target state.
In \fref{fig:number5} we present the pins showing the positions and the absolute values of the coefficients of  the $N=9$ constituent coherent states in the coherent-state superposition approximating the number state $\ket{3}$  on a lattice and the Wigner function of the approximating state which is almost identical with the one of the target state.

For the number states we have found that high accuracy can be achieved not only at the optimal value of $d$ but in a range of distances around this value.
\begin{figure}
	\centerline{\includegraphics[width=0.8\columnwidth]{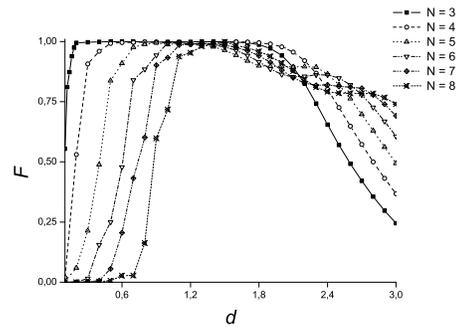}}
	\caption{\label{fig:numberstates}Precisions of approximating number states $\ket{N}$ on a lattice depending on the parameter $d$.}
\end{figure}
This finding is shown in \fref{fig:numberstates} where the quantity $F=\sqrt{1-\epsilon^{(\rm lattice)}}$ is presented as a function of $n$. The quantity $F$ is the absolute value of the scalar product $|\braket{\psi_N}{\Psi}|$ sometimes used as a form of fidelity in the literature. The figure shows that as the photon number $n$ increases, the width of the acceptable range of high accuracy decreases. The lower limit of this range shifts to the direction of larger distances while the upper limit shows a more complicated behavior, but the accuracy of the approximation relevantly decreases for distances $d>1.7$.

\begin{table}
\caption{\label{tab:numsup}
The misfits $\epsilon^{(\rm lattice)}$ and $\epsilon^{(\rm circle)}$ of the approximation  of number-state superpositions on a lattice and on a circle with 9 coherent states and the optimal distances $d_{\rm opt}$ and radii $R_{\rm opt}$ for increasing number of superposed states.} 
\begin{indented}
\item[]\begin{tabular}{lllll}
		\br
			State & $\epsilon^{(\rm lattice)}$ &$d_{\rm opt}$ & $\epsilon^{(\rm circle)}$ & $R_{\rm opt}$ \\
		\mr
			$\Psi_{01}$& $2 \times 10^{-6}$  & 0.257 & $2.2 \times 10^{-5}$ & 0.54\\
			$\Psi_{012}$ & $1.8 \times 10^{-5}$ &0.53 & $4\times10^{-4}$& 0.9\\
			$\Psi_{0123}$& $6.6 \times 10^{-5}$ & 0.79 & 0.0016 & 1.17\\
			$\Psi_{01234}$ & 0.0017 &1.37 & 0.0061 & 1.32 \\
			$\tilde{\Psi}_{01234}$ &  0.0016 & 1.33 & 0.0035 & 1.15 \\
		\br
		\end{tabular}
\end{indented}		
\end{table}

Let us now consider the approximation of the number state superpositions
\begin{eqnarray}
\Psi_{01}=\frac{1}{\sqrt{2}}(\ket{0}+\ket{1}),\\
\Psi_{012}=\frac{1}{\sqrt{3}}(\ket{0}+ \ket{1} + \ket{2}),\\
\Psi_{0123}=\frac{1}{2}(\ket{0}+ \ket{1} + \ket{2} + \ket{3}),\\
\Psi_{01234}=\frac{1}{\sqrt{5}}(\ket{0}+ \ket{1} + \ket{2} + \ket{3} + \ket{4}),\\
\tilde{\Psi}_{01234}=\frac{1}{\sqrt{72}}(7 \ket{0} + 3 \ket{1} + 2\ket{2} + \ket{3} + 3\ket{4})
\end{eqnarray}
by lattice and circle coherent-state superpositions. In \tref{tab:numsup} we present the misfits $\epsilon^{(\rm lattice)}$ and $\epsilon^{(\rm circle)}$ of the approximations and the optimal distances $d_{\rm opt}$ of adjacent lattice elements, and the optimal radii $R_{\rm opt}$. The table shows that the accuracy of the approximation on a lattice is quite high for all the considered superpositions, and it is always higher than the accuracy of the approximation on a circle. Increasing the number of constituent number states in the target state including higher number states results in larger optimal distances $d_{\rm opt}$ and optimal radii $R_{\rm opt}$. We note that generation of such types of superpositions can be important for some quantum information protocols \cite{Ralph2001, Adesso2009, Marek2009}. Several schemes have been developed for producing such states in the literature \cite{Pegg1998, Marek2009, Lee2010}. From the results presented here it can be seen that the construction and generation of these states by coherent-state superposition on a lattice can be a rather effective method.

\begin{figure}
\centering
\includegraphics[width=0.8\columnwidth]{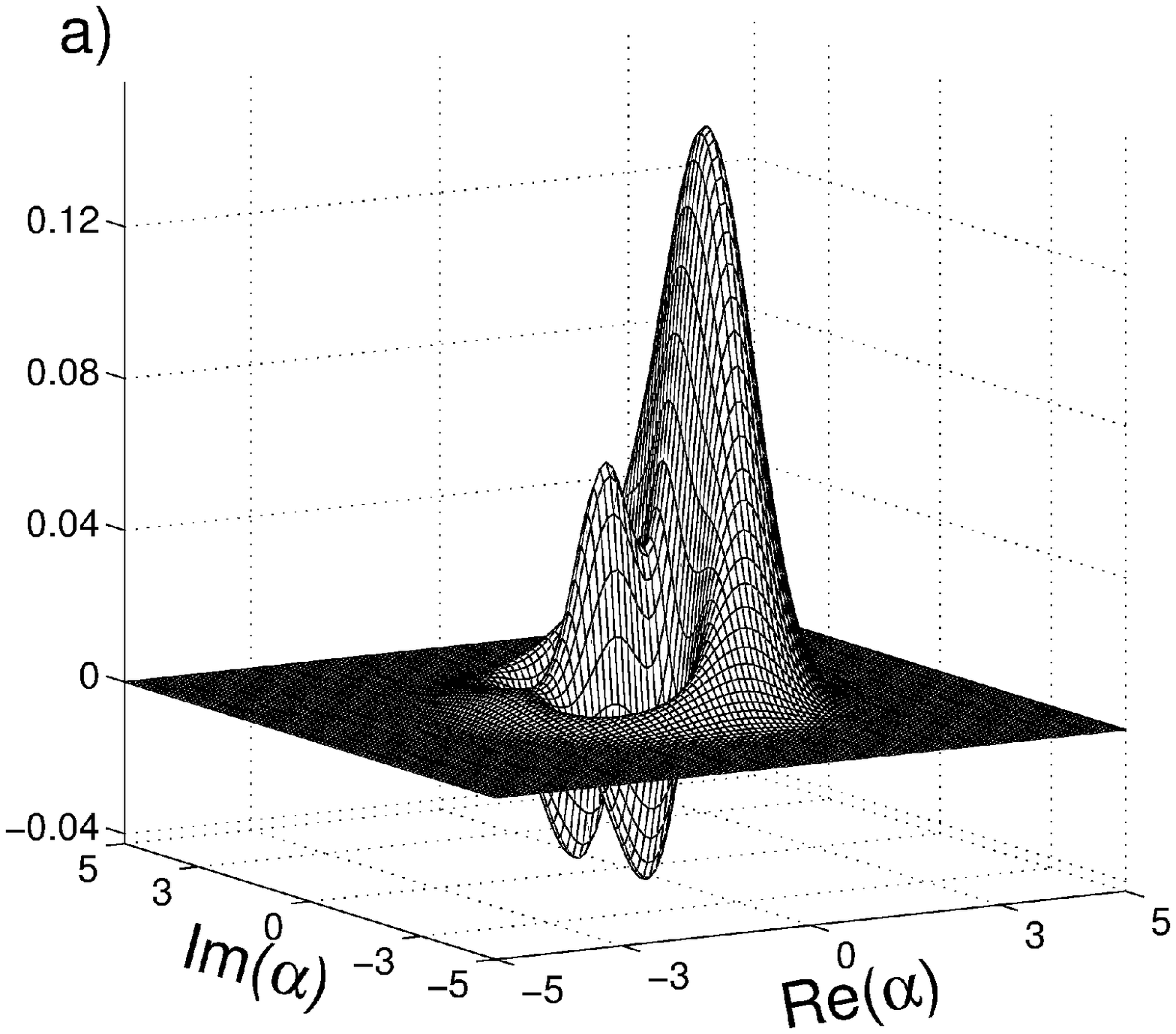}
\includegraphics[width=0.8\columnwidth]{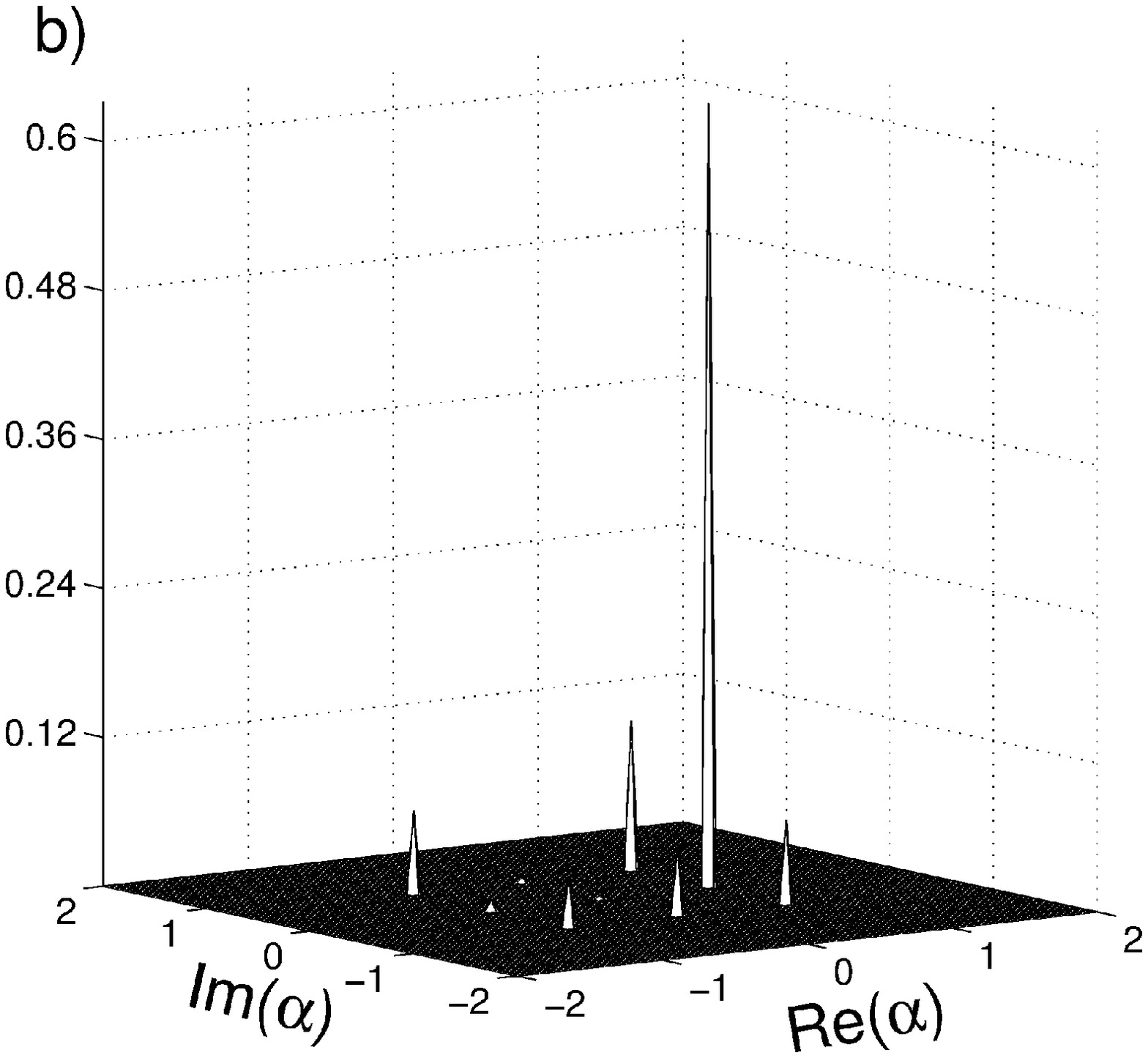}
\caption{\label{fig:number123}Number state superposition $\frac{1}{\sqrt{3}}(\ket{0}+\ket{1}+\ket{2}) $ approximated by $N=9$ coherent states on a lattice (a) Wigner function $W(\alpha)$ of the approximating coherent-state superposition, (b) pins showing the positions and the absolute values of the coefficients of the constituent coherent states.}
\end{figure}

In \fref{fig:number123} we present the pins showing the positions and the absolute values of the coefficients of  the $N=9$ constituent coherent states in the coherent-state superposition approximating the $\frac{1}{\sqrt{3}}(\ket{0}+\ket{1}+\ket{2})$ state on a lattice and the Wigner function of the approximating state.
This function is practically perfect showing the high accuracy of the approximation.

Next we consider the approximation of squeezed number states. \Tref{tab:sn1rot} presents the misfits $\epsilon^{(\rm lattice)}$ and $\epsilon^{(\rm circle)}$ of the approximations, the optimal distances $d_{\rm opt}$ of adjacent lattice elements, and  radii $R_{\rm opt}$ for various phases $\theta$ and for a constant magnitude $r=0.5$ of squeezing and a constant photon number $n=1$. The data show that the accuracy of the approximation is generally good but really high around $\theta=0$, $\theta=\pi$, and $\theta=\pi/2$ with the range of $|\Delta \theta|\lesssim\pi/12$. In this case the state is elongated in the direction of the real axis, the imaginary axis, and the diagonal of the lattice, respectively. In these cases the approximation on a lattice is better than the one on a circle. Though there is no general rule in the change of the optimal distance $d_{\rm opt}$ and radius $R_{\rm opt}$, one can observe that the optimal distances are smaller for the states with phases of squeezing around $\pi/2$ ($\theta=\pi/2\pm\pi/6$).
\begin{table}
	\caption{\label{tab:sn1rot}
The misfits $\epsilon^{(\rm lattice)}$ and $\epsilon^{(\rm circle)}$ of the approximation  of squeezed number states on a lattice and on a circle with 9 coherent states and the optimal distances $d_{\rm opt}$ and radii $R_{\rm opt}$ for various phase $\theta$ and a constant magnitude $r=0.5$ of squeezing and a constant photon number $n=1$.}
\begin{indented}
\item[]\begin{tabular}{lllll}
	\br
		State & $\epsilon^{(\rm lattice)}$ &$d_{\rm opt}$ & $\epsilon^{(\rm circle)}$ & $R_{\rm opt}$\\
	\mr
		$\ket{1,0.5,0} $ & 0.0017  & 1.44& 0.0019 & 1.53 \\
		$\ket{1,0.5\exp(\rmi \pi/12),0} $ & 0.0069   & 1.34&0.0034 &1.53 \\
		$\ket{1,0.5\exp(\rmi \pi/6),0}$ & 0.008 &1.15&0.0012 & 1.63\\
		$\ket{1,0.5\exp(\rmi \pi/3),0}$ & 0.0065  & 1.14&0.004 &1.49 \\
		$\ket{1,0.5\exp(\rmi 5\pi/12),0}$ & 0.0025  & 1.13&0.0049 & 1.44\\
		$\ket{1,0.5\exp(\rmi \pi/2),0}$ & 0.0033  & 1.04&0.0042 & 1.39\\
		$\ket{1,0.5\exp(\rmi 7\pi/12),0}$ & 0.0032  & 1.07&0.0063 &1.43 \\
		$\ket{1,0.5\exp(\rmi 2\pi/3),0}$ & 0.006  & 1.13&0.0044 & 1.42\\
		$\ket{1,0.5\exp(\rmi 5\pi/6),0}$ & 0.0074 & 1.1&0.005 & 1.51\\
		$\ket{1,0.5\exp(\rmi 11\pi/12),0} $ & 0.007  & 1.35&0.0032 & 1.54 \\
		$\ket{1,0.5\exp(\rmi \pi),0}$ & 0.0018& 1.44  & 0.0029& 1.48\\
	\br
	\end{tabular}
\end{indented}
\end{table}
\begin{figure}
\centering
\includegraphics[width=0.8\columnwidth]{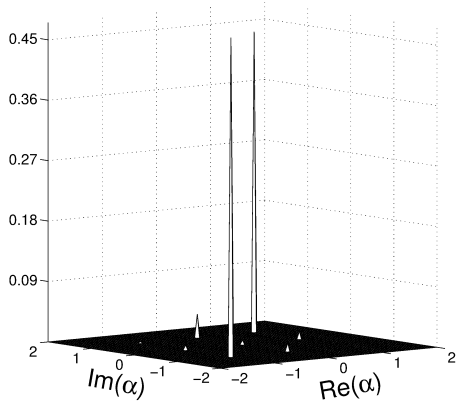}
\caption{\label{fig:tuskesnl} Pins showing the positions and the absolute values of the coefficients of the constituent coherent states for squeezed number state $\ket{1,0.5\exp(\rmi \pi/2),0}$ with $N=9$ coherent states on a lattice.}
\end{figure}
In \fref{fig:tuskesnl} we show the pins representing the positions and the absolute values of the coefficients of the constituent coherent states for squeezed number state $\ket{1,0.5\exp(\rmi \pi/2),0}$ with $N=9$ coherent states on a lattice.

Finally, let us consider amplitude-squeezed states
\begin{equation}
\ket{\Psi_A(u,\delta,R)}=\mathcal{N}\int\limits_{-\pi}^{\pi}\exp\left(-\frac{1}{2}u^2\phi^2-\rmi\delta\phi\right)\ket{R\rme^{\rmi\phi}}{\rm d}\phi\label{eq:5:asstate}
\end{equation}
defined by Gaussian continuous coherent-state superpositions in phase space. $\mathcal{N}$ is a normalization constant.

For calculational purposes it is convenient to use the number-state representation of the states:
\begin{equation}
\ket{\Psi_A(u,\delta,R)}=\sum_{n=0}^{\infty}C_n\ket{n},
\label{eq:5:asstate:pnex}
\end{equation}
where
\begin{equation}
C_n=\braket{n}{\Psi_A(u,\delta,R)}=\mathcal{N}\frac{2\pi R^n}{u\sqrt{n!}}\exp\left[-\frac{(\delta-u)^2}{2u^2}\right].
\label{eq:asstate_photon}
\end{equation}
These states tend to the coherent state $R$ in the limit $u\to\infty$ and yield the photon number state $\delta$ when $\delta$ is a nonnegative integer in the limit $u\to0$, while $\delta=R^2$, the mean values of the photon number for these limiting states are equal.

In \tref{tab:as1} the misfits $\epsilon^{(\rm lattice)}$ and $\epsilon^{(\rm circle)}$ of the approximations of various amplitude-squeezed states on a lattice and on a circle and the optimal parameters $d_{\rm opt}$ and $R_{\rm opt}$ are presented.
The table shows that the accuracy of the approximation on both formations is high for all the considered states. The accuracy on a circle is always higher than on a lattice except for the state $\ket{\Psi_A(2,4,2)}$. The optimal distance $d_{\rm opt}$ of the lattice is growing with increasing values of the parameter $u$.
\begin{table}
	\caption{\label{tab:as1}
The misfits $\epsilon^{(\rm lattice)}$ and $\epsilon^{(\rm circle)}$ of the approximation  of amplitude-squeezed states on a lattice and on a circle with 9 coherent states and the optimal distances $d_{\rm opt}$ and radii $R_{\rm opt}$ for various values of $u$ and fixed parameters $\delta=4$ and $R=2$.}
\begin{indented}
\item[]\begin{tabular}{lllll}
		\br
			State & $\epsilon^{(\rm lattice)}$ &$d_{\rm opt}$ & $\epsilon^{(\rm circle)}$ & $R_{\rm opt}$\\
		\mr
			$\ket{\Psi_A(0.5,4,2)} $ & 0.003  & 0.9& $5 \times 10^{-5}$ & 1.65\\
			$\ket{\Psi_A(1,4,2)}  $ & 0.0048 &0.98& $2\times10^{-4}$& 1.74\\
			$\ket{\Psi_A(1.5,4,2)}  $ & 0.0066 & 1.15 & $4\times10^{-4}$& 1.79\\
			$\ket{\Psi_A(1.6,4,2)}  $ & 0.0093 & 1.16 & $2.2\times10^{-4}$& 1.72\\
			$\ket{\Psi_A(1.7,4,2)}  $ & 0.0058 & 1.16 & $9\times10^{-4}$ & 1.87\\
			$\ket{\Psi_A(1.9,4,2)}  $ & 0.0142 & 1.51 & 0.01 & 1.84\\
			$\ket{\Psi_A(2,4,2)} $ & 0.0105 & 1.41 & 0.012 & 1.86\\
			$\ket{\Psi_A(3,4,2)}  $ & 0.011 &1.9  & 0.0013 & 1.78\\
			$\ket{\Psi_A(5,4,2)}  $ & 0.0031 & 1.95 & $7\times10^{-4}$ & 1.93\\
			$\ket{\Psi_A(10,4,2)}  $ & $3\times10^{-4}$ & 1.99 & $1.3\times10^{-4}$& 1.98\\
		\br
		\end{tabular}
		\end{indented}
\end{table}

\begin{figure}
\centering
\includegraphics[width=0.8\columnwidth]{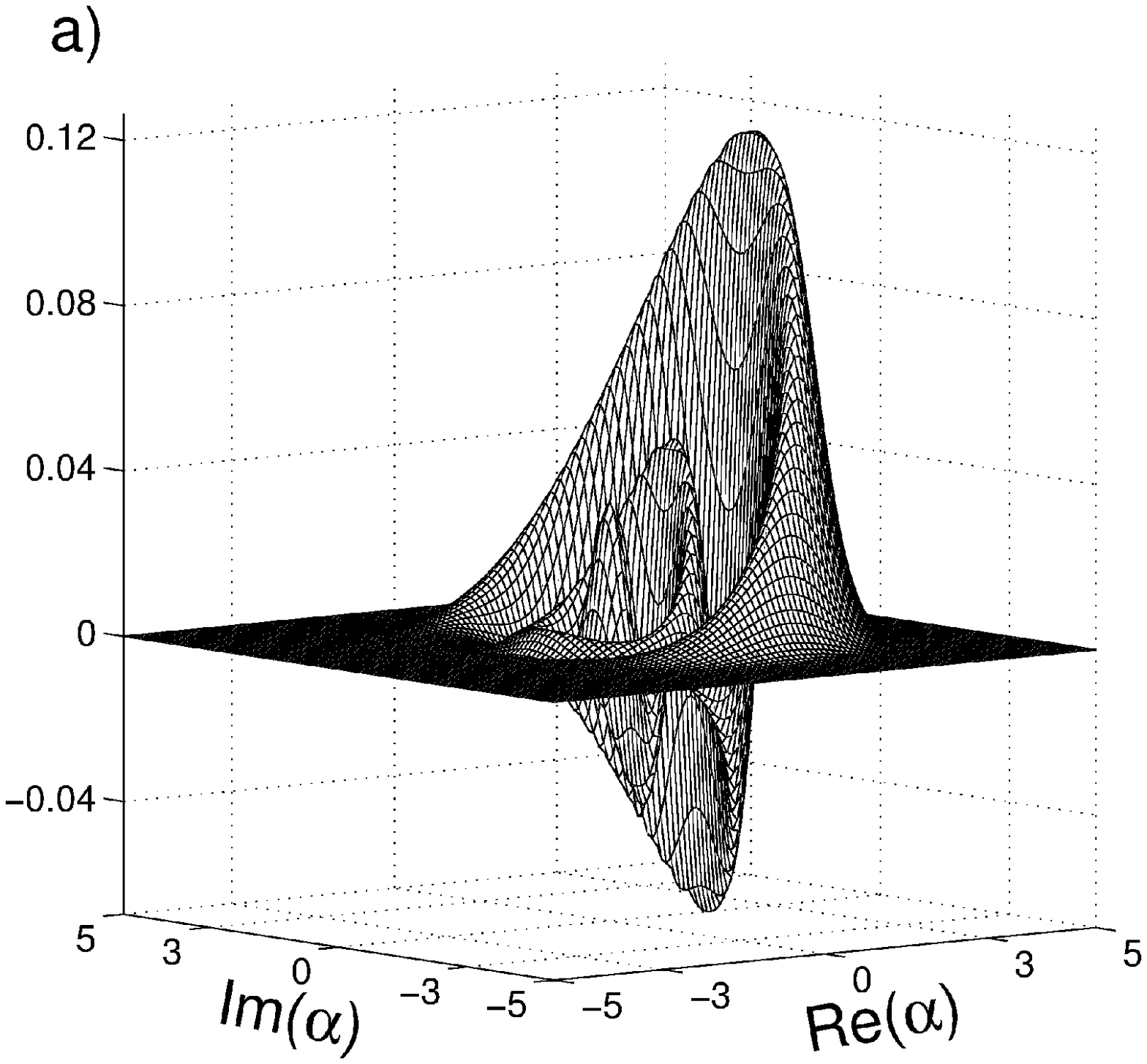}
\includegraphics[width=0.8\columnwidth]{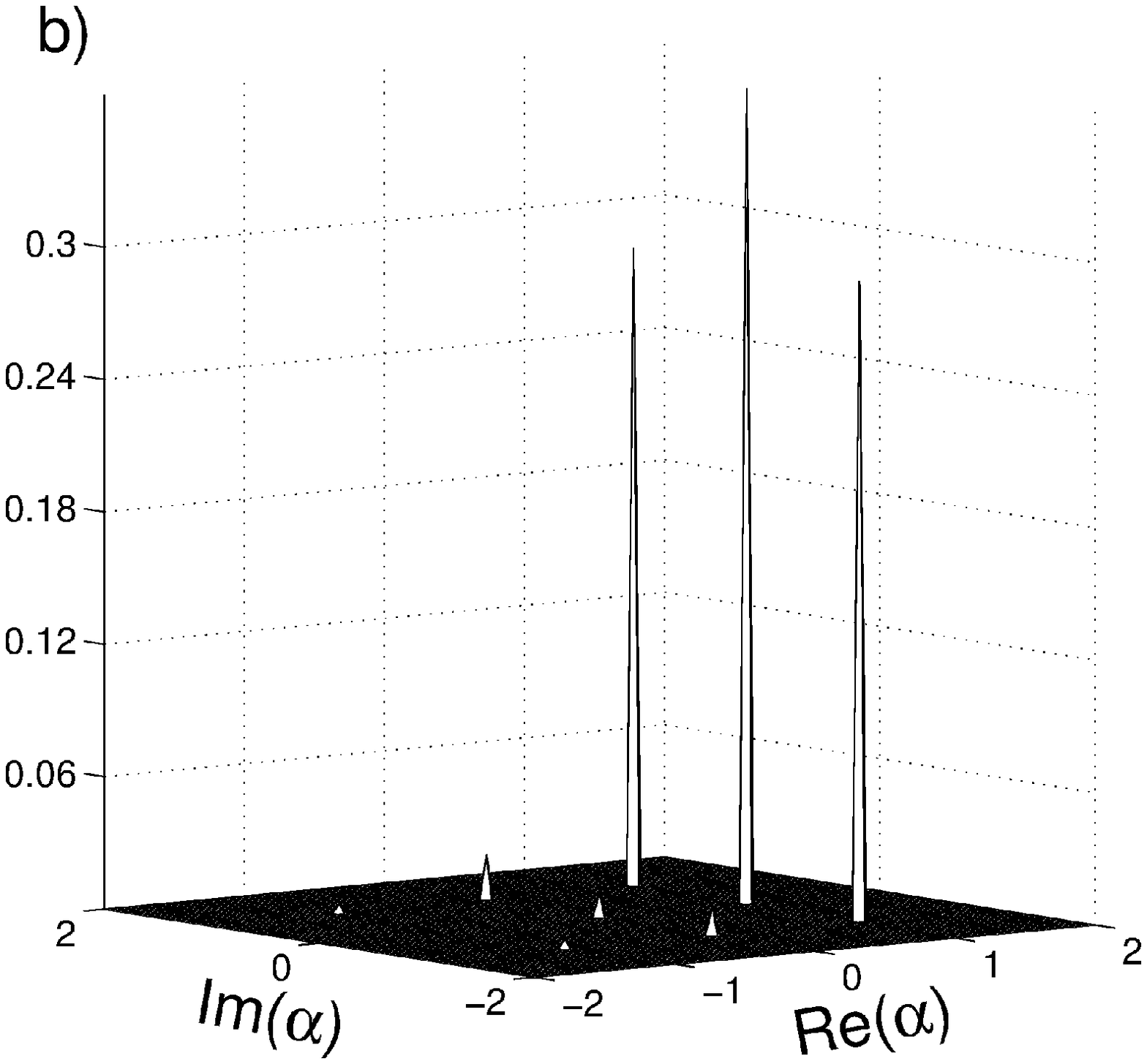}
\caption{\label{fig:banan}Amplitude-squeezed state $\ket{\Psi_A(1,4,2)}$ approximated by $N=9$ coherent states on a lattice (a) Wigner function $W(\alpha)$ of the approximating coherent-state superposition, (b) pins showing the positions and the absolute values of the coefficients of the constituent coherent states.}
\end{figure}

In \fref{fig:banan} we present the pins showing the positions and the absolute values of the coefficients of  the $N=9$ constituent coherent states in the coherent-state superposition approximating the amplitude-squeezed state $\ket{\psi_A(1,4,2)}$ on a lattice and the Wigner function of the approximating state. In this case the Wigner function is also practically perfect.

From these results one can conclude that lattice coherent-state superpositions are universal for approximating nonclassical states of harmonic oscillator systems. The accuracy of the approximation is high for several quantum states. Higher accuracy can be obtained by circle superpositions only for quantum states having circular symmetry in the Wigner function \cite{Szabo1996}.

\section{Conclusions}

We have considered the optimal approximation of certain quantum states
of a harmonic oscillator with the superposition of a finite number of
coherent states on an ellipse and on a lattice in phase space. All the coherent-state superpositions we have
considered are feasible in current experiments.  We have optimized
numerically the parameters of the chosen
geometry and the coefficients of the coherent states via a genetic algorithm, in order to obtain
the best feasible approximation.

First, we placed the coherent states equidistantly in their phase on an
ellipse.  As the ellipse fits to the shape of the Wigner function of
certain states, we have expected that for such states we can obtain an
approximation superior in quality to the one based on the
one-dimensional coherent-state representation of the state, that is,
constructed from states along a line in phase space. We have found
this intuition to be appropriate. In particular, for squeezed number
states in a certain parameter range, the elliptical approximation
outperforms the one on a line.

Next we have considered a $3\times 3$ equidistant lattice around the
origin of phase space. These are the kinds of superpositions which
seem to be feasible in travelling-wave optical experiments. A
relatively large set of states, including special number-state
superpositions, appear to be better approximated in this geometry than
with a superposition on a circle.

\ack
We are sincerely grateful to Margarita A. Man'ko and Vladimir I. Man'ko for their long-standing collaboration with our group.
We are grateful for the support of the Hungarian Scientific Research Fund OTKA
(Contract No.\ K83858). We thank M. Koniorczyk for useful discussions.

\end{document}